\pdfoutput=1

\documentclass[a4paper]{jpconf}
\usepackage{graphicx}
\usepackage{microtype}
\usepackage[colorlinks=true,linkcolor=magenta,citecolor=magenta]{hyperref}

\newcommand{\bb}{\ensuremath{\beta\beta}}
\newcommand{\bbonu}{\ensuremath{\beta\beta0\nu}}
\newcommand{\bbtnu}{\ensuremath{\beta\beta2\nu}}
\newcommand{\mbb}{\ensuremath{m_{\beta\beta}}}
\newcommand{\Qbb}{\ensuremath{Q_{\beta\beta}}}
\newcommand{\XE}{\ensuremath{^{136}\mathrm{Xe}}}
\newcommand{\BI}{\ensuremath{^{214}\mathrm{Bi}}}
\newcommand{\TL}{\ensuremath{^{208}\mathrm{Tl}}}
\newcommand{\ckky}{\ensuremath{\mathrm{counts/(keV~kg~year})}}

\begin{document}

\title{Status and physics potential of NEXT-100}

\author{J.~Mart\'in-Albo and J.J.~G\'omez-Cadenas,\\ on behalf of the NEXT Collaboration}
\address{Instituto de F\'isica Corpuscular (IFIC), CSIC \& Universidad de Valencia\\
Calle Catedr\'atico Jos\'e Beltr\'an, 2, 46980 Paterna, Valencia, Spain}
\ead{justo.martin-albo@ific.uv.es, gomez@mail.cern.ch}

\begin{abstract}
The NEXT-100 time projection chamber, currently under construction, will search for neutrinoless double beta decay (\bbonu) using 100--150 kg of high-pressure xenon gas enriched in the \XE\ isotope to $\sim91\%$. The detector possesses two important features for \bbonu\ searches: very good energy resolution (better than 1\% FWHM at the $Q$ value of \XE) and event topological information for the distinction between signal and background. Furthermore, the technique can be extrapolated to the ton-scale, thus allowing the full exploration of the inverted hierarchy of neutrino masses.
\end{abstract}

\section{Introduction} \label{sec:Introduction}
Neutrinoless double beta decay (\bbonu) is a postulated very slow nuclear transition in which two neutrons undergo $\beta$ decay simultaneously and without the emission of neutrinos \cite{GomezCadenas:2011it, Avignone:2007fu}. The importance of this process can hardly be overstated. An unequivocal observation would establish that massive neutrinos are Majorana particles \cite{Furry:1939qr, Schechter:1980gr} ---~that is, identical to their antiparticles~---, implying that a new physics scale beyond the Standard Model must exist \cite{Beringer:1900zz, GonzalezGarcia:2007ib}. Furthermore, it would prove that total lepton number is not conserved, suggesting that this new physics could also be the reason for the observed asymmetry between matter and antimatter in the universe~\cite{Fukugita:1986hr, Davidson:2008bu}.

Many mediating mechanisms ---~involving, in general, physics beyond the Standard Model~--- have been proposed for \bbonu, the simplest one being the virtual exchange of light Majorana neutrinos \cite{GomezCadenas:2011it}. Assuming this to be the dominant mechanism at low energies, the decay rate of \bbonu\ can be written as \cite{Avignone:2007fu}
\begin{equation}
(T_{1/2}^{0\nu})^{-1} = \frac{F_{N}}{m_{e}}\, \mbb^{2}\, .
\end{equation}
In this equation, $F_{N}$ is a nuclear structure factor that depends on the particular isotope that is disintegrating, $m_{e}$ is the mass of the electron, and \mbb\ is the \emph{effective Majorana mass} of the electron neutrino:
\begin{equation}
\mbb = \left|\sum_{i} U^{2}_{ei}\ m_{i}\right| ,
\end{equation}
where $m_{i}$ are the neutrino mass eigenstates and $U_{ei}$ are elements of the neutrino mixing matrix. In consequence, a measurement of the \bbonu\ decay rate would provide direct information on neutrino masses \cite{GomezCadenas:2011it}.

\section{Experimental searches for neutrinoless double beta decay} \label{sec:ExperimentalSearches}
The detectors used to search for \bbonu\ are designed, in general, to measure the energy of the radiation emitted by a \bbonu\ source. In a neutrinoless double beta decay, the sum of the kinetic energies of the two released electrons is always the same, and is equal to the mass difference between the parent and the daughter nuclei: $\Qbb \equiv M(Z,A)-M(Z+2,A)$. However, due to the finite energy resolution of any detector, \bbonu\ events would be reconstructed within a given energy range centered around \Qbb\ and typically following a gaussian distribution. Other processes occurring in the detector can fall in that region of energies, thus becoming a background and compromising drastically the sensitivity of the experiment \cite{GomezCadenas:2010gs}.

All double beta decay experiments have to deal with an intrinsic background, the standard two-neutrino double beta decay (\bbtnu), that can only be suppressed by means of good energy resolution. Backgrounds of cosmogenic origin force the underground operation of the detectors. Natural radioactivity emanating from the detector materials and surroundings can easily overwhelm the signal peak, and hence careful selection of radiopure materials is essential. Additional experimental signatures that allow the distinction of signal and background are a bonus to provide a robust result.

Besides energy resolution and control of backgrounds, several other factors such as detection efficiency and scalability to large masses must be taken into consideration in the design of a double beta decay experiment. The simultaneous optimization of all these parameters is most of the time conflicting, if not impossible, and consequently many different experimental techniques have been proposed. In order to compare them, a figure of merit, the experimental sensitivity to \mbb, is normally used \cite{GomezCadenas:2010gs}:
\begin{equation}
\mbb \propto \sqrt{1/\varepsilon}\, \left(\frac{b\ \delta E}{M\ t} \right)^{1/4}, \label{eq:sensi}
\end{equation}
where $\varepsilon$ is the signal detection efficiency, $M$ is the \bb\ isotope mass used in the experiment, $t$ is the data-taking time, $\delta E$ is the energy resolution and $b$ is the specific background rate in the region of interest around \Qbb\ (expressed in counts per kg of \bb\ isotope, year and keV).

Until very recently, the experimental searches for \bbonu\ had been dominated by germanium calorimeters, mainly due to their excellent energy resolution. In particular, for about a decade the best limit to the half-life of \bbonu\ was the one set by the Heidelberg-Moscow (HM) experiment: $T_{1/2}^{0\nu}(^{76}\mathrm{Ge}) \geq 1.9\times10^{25}$ years at 90\% CL \cite{KlapdorKleingrothaus:2000sn}. A subgroup of this experiment interpreted the same data as evidence of a positive signal, with a best value for the half-life of $1.5\times10^{25}$~years, or 0.39~eV in terms of \mbb\ \cite{KlapdorKleingrothaus:2001ke}. This claim 
has been practically ruled out by two new experiments searching for \bbonu\ in \XE, EXO-200 \cite{Auger:2012gs, Auger:2012ar} and KamLAND-Zen \cite{KamLANDZen:2012aa, Gando:2012zm}, that published their first results during 2012. Their basic experimental parameters, as defined in equation~(\ref{eq:sensi}), are collected in Table~\ref{tab:ExpParams}. The combination of the limits reported by the two experiments gives $T_{1/2}^{0\nu}(\XE) > 3.4\times10^{25}$~years (90\%~CL), refuting the HM claim at $>97.5\%$~CL \cite{Gando:2012zm}. 

Xenon is indeed an interesting species for double beta decay searches. Two of its natural isotopes, $^{134}$Xe and \XE, are \bbonu-decaying candidates. The latter, having a higher \Qbb\ value (2458 keV \cite{McCowan:2010zz}), is preferred for \bbonu\ searches because the decay rate is proportional to $\Qbb^{5}$ and the radioactive backgrounds are less abundant at higher energies. The \bbtnu\ mode of \XE\ ($\sim2.3\times10^{21}$~years) is slow compared to that of other sources, and therefore the experimental requirement for energy resolution is less stringent. Moreover, the process of isotopic enrichment is relatively simple and less costly than for the other \bbonu\ isotopes, and consequently \XE-based experiments are the most obvious candidates for a future multi-ton experiment.

The KamLAND-Zen experiment is a modification of the well-known KamLAND neutrino detector \cite{Abe:2009aa}. A transparent balloon, $\sim3$~m diameter, containing 13 tons of liquid scintillator loaded with 330~kg of enriched xenon (91\% of \XE) is suspended at the center of KamLAND. The scintillation light generated by events occurring in the detector is recorded by an array of photomultipliers surrounding it. From the detected light pattern, the position of the event vertex is reconstructed with a spatial resolution of about $15~\mathrm{cm}/\sqrt{E(\mathrm{MeV})}$. The energy resolution is $(6.6\pm0.3)\%/\sqrt{E(\mathrm{MeV})}$, that is, 9.9\% FWHM at the $Q$ value of \XE. The signal detection efficiency is 42\% due to the tight fiducial cut introduced to reject backgrounds originating in the balloon. The achieved background rate in the energy window between 2.2~MeV and 3.0~MeV is $0.001~\ckky$. The dominant source of background is $^{100m}$Ag, a $\beta^{-}$ emitter. Its presence in the detector may be due to contamination of the balloon by Fukushima fallout during fabrication or cosmogenic production by Xe spallation.

EXO-200, unlike KamLAND-Zen, uses the enriched xenon as both source and detection medium. The detector is a cylindrical time projection chamber (TPC) filled with 110 kg of liquid xenon (LXe) enriched to 81\% in \XE. The interaction of charged particles with the LXe produces both a scintillation and an ionization signal. The TPC is divided into two symmetric halves separated by a cathode grid, and each end is instrumented with a pair of crossed wire planes that detect the ionization, and behind those, an array of 250 APDs is used to record the scintillation light. The energy resolution obtained combining the two signals is 3.9\% FWHM at \Qbb. The topological reconstruction of the events is used for discriminating signal and background. The background rate in the region of interest around \Qbb\ is $1.5\times10^{-3}$~\ckky. The dominant sources of backgrounds are the external high-energy gammas and the radioactive contaminants in the detector vessel.

\begin{table}
\lineup
\centering
\caption{Basic experimental parameters of the three \XE-based double beta decay experiments: mass of \XE, $M$; signal detection efficiency, $\varepsilon$; energy resolution at the $Q$ value of \XE, $\delta E$; and background rate, $b$, in the region of interest around \Qbb\ expressed in \ckky\ (shortened as ckky). } \label{tab:ExpParams}
\begin{tabular}{lcccc}
\br
Experiment & $M$ (kg) & $\varepsilon$ (\%) & $\delta E$ (\% FWHM) & $b$ ($10^{-3}$~ckky) \\
\mr
EXO-200 		& \090  & 0.62 & 3.9 & 1.5 \\
KamLAND-Zen & 300 & 0.42 & 9.9 & 1.0 \\
NEXT-100 	& \090  & 0.25 & 0.7 & 0.5 \\
\br
\end{tabular}
\end{table}

\section{NEXT-100, high-pressure xenon gas for \boldmath{\bbonu} searches} \label{sec:NextConcept}
The NEXT-100\footnote{NEXT stands for \emph{Neutrino Experiment with a Xenon gas Time projection chamber}.} detector \cite{NEXT:2012haa} will search for the neutrinoless double beta decay of \XE\ using a time projection chamber filled with 100--150 kg of enriched xenon gas at 10--15 bar pressure. Such a detector offers both very good energy resolution (better than 1\% FWHM at 2.5 MeV) and event topological information to discriminate between signal and background. As we will see, this combination results in excellent sensitivity to \bbonu. In addition, the technology can be extrapolated to a ton-scale experiment, thus allowing the full exploration of the inverted hierarchy of neutrino masses \cite{GomezCadenas:2011it}. The detector is currently under construction, and its installation and commissioning at the \emph{Laboratorio Subterr\'aneo de Canfranc} (LSC), in Spain, are planned for the first semester of 2014.

During the last three years, the NEXT Collaboration has developed an R\&D program with the specific goal of proving the performance of the technology. This program has resulted in the construction and operation of the NEXT-DEMO \cite{Alvarez:2012np} and NEXT-DBDM \cite{Alvarez:2012hh} prototypes. The first results of these detectors have been presented at this conference \cite{Oliveira:2013}.

Gaseous xenon, as a detection medium, provides, like LXe, \emph{scintillation} and \emph{ionization} as primary signals. The former is used in NEXT to establish the start-of-event time ($t_{0}$), while the latter is used for calorimetry and tracking. In its gaseous phase, xenon can provide high energy resolution, in principle as good as 0.3\% FHWM at the $Q$ value of \XE\ \cite{Nygren:2009zz}. In order to achieve optimal energy resolution, the ionization signal is amplified in NEXT using the electroluminescence (EL) of xenon: the electrons liberated by ionizing particles passing through the gas are first drifted towards the TPC anode by a weak electric field ($\sim300$~V/cm), entering then into another region where they are accelerated by a high electric field ($\sim25$~kV/cm at 10 bar), intense enough so that the electrons can excite the xenon atoms but not enough to ionize them. This excitation energy is ultimately released in the form of proportional (with sub-poissonian fluctuations) secondary scintillation light.

NEXT-100 will have different readout systems for calorimetry and tracking. An array of 60 photomultiplier tubes (Hamamatsu R11410-10 PMTs), the so-called \emph{energy plane}, located behind the TPC cathode detects a fraction of the secondary scintillation light to provide a precise measurement of the total energy deposited in the gas. These PMTs detect as well the primary scintillation, used to signal the start of the event. The forward-going secondary scintillation is detected by a dense array of 1-mm$^{2}$ silicon photomultipliers (SiPMs), known as the \emph{tracking plane}, located behind the anode, very close to the EL region, and is used for event topological reconstruction.

\begin{figure}
\centering
\includegraphics[height=5cm]{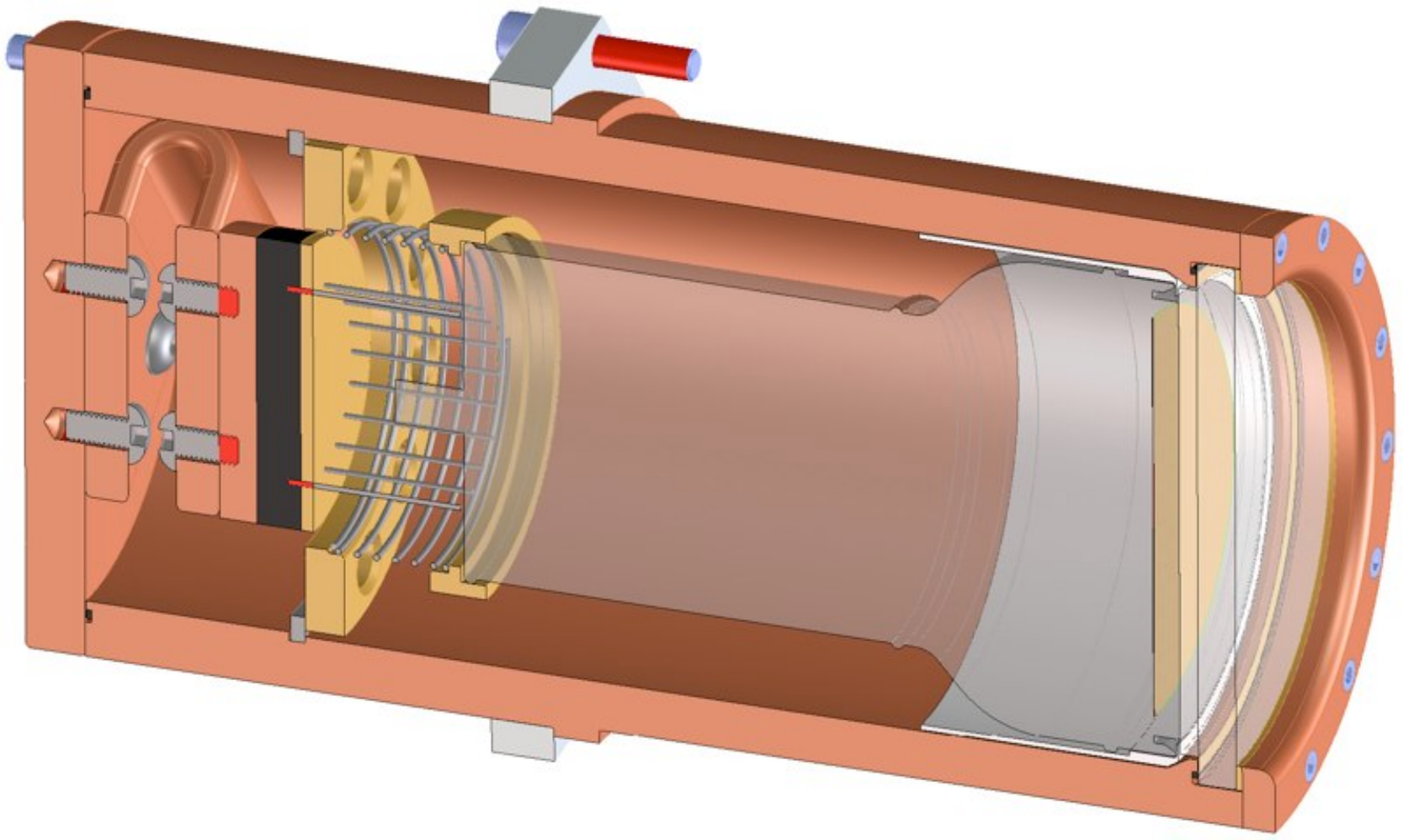}
\includegraphics[height=5cm]{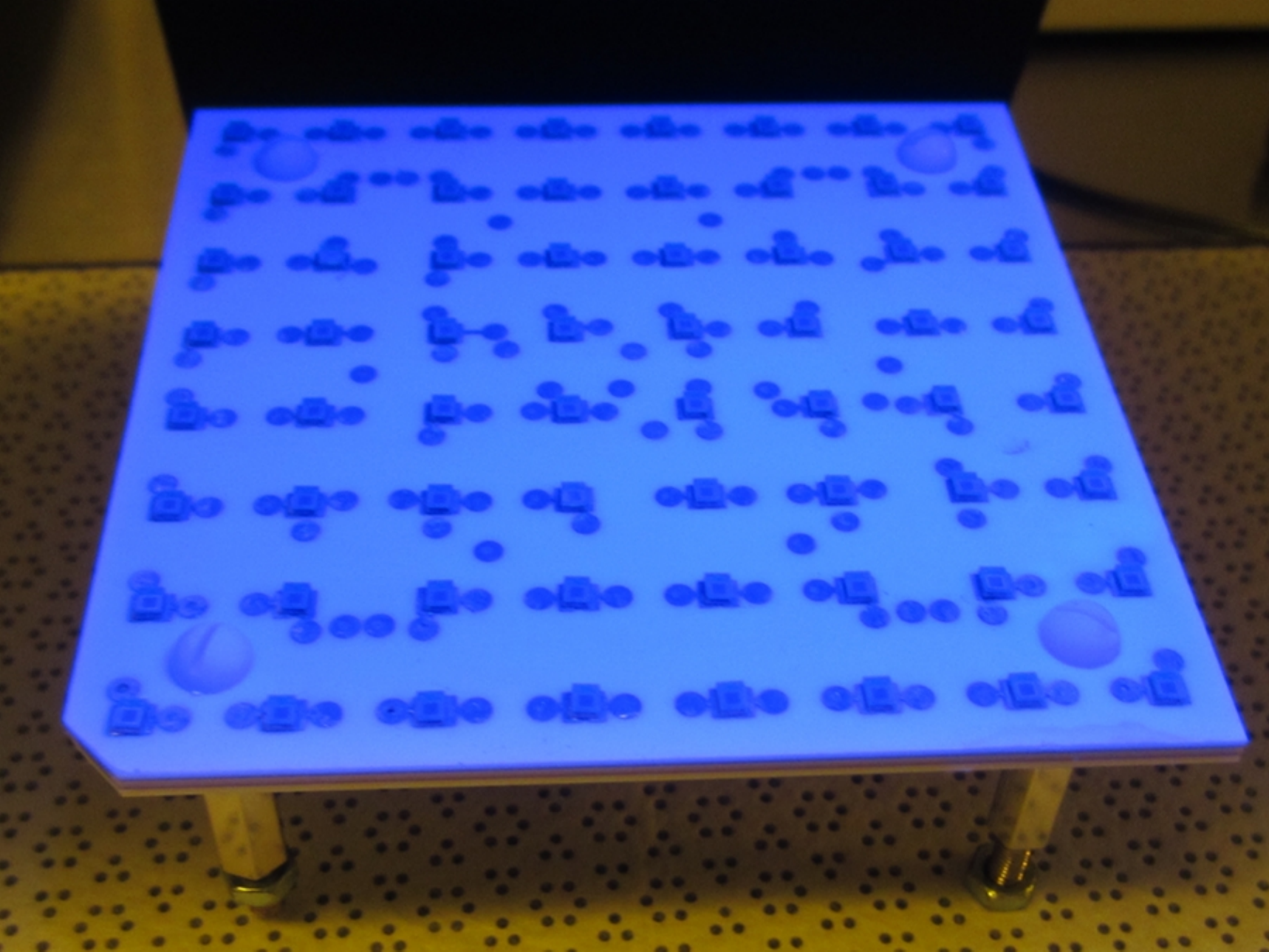}
\caption{Left: Drawing of a Hamamatsu R11410-10 PMT inside its pressure-resistant copper enclosure. Right: A \emph{dice board} containing 64 ($8\times8$) SiPMs illuminated with a UV lamp after being coated with TPB.} \label{fig:Sensors}
\end{figure}

The R11410-10 PMT was specially developed for radiopure, xenon-based detectors. However, pressure-resistance tests run by the manufacturer showed that this PMT cannot withstand pressures above 6 atmospheres. Therefore in NEXT-100 they will be sealed into individual pressure resistant, vacuum tight copper enclosures coupled to sapphire windows (see figure~\ref{fig:Sensors}, left). The chosen SiPM for NEXT-100 is the S10362-11-050P model by Hamamatsu. This device has an active area of 1 mm$^{2}$, 400 sensitive cells (50 $\mu$m size) and high photon detection efficiency in the blue region (about 50\% at 440 nm). The dark count rate is 0.4~MHz, that is, less than 1 event per microsecond (which is the considered sampling time). This random noise events have amplitudes of up to 8 photoelectrons, and thus a digital threshold at those levels should lead to an insignificant noise rate in NEXT-100 without affecting the tracking performance. The SiPMs will be mounted in cuflon or cirlex boards (depending on the radiopurity of these materials) spaced 1 cm (see figure~\ref{fig:Sensors}, right). The photon detection efficiency (PDE) of the chosen SiPMs peaks in the blue region of the spectrum, and they are not sensitive below 200~nm, where the emission spectrum of xenon lies. Consequently, the boards will be coated with tetraphenyl butadiene (TPB), a wavelength shifter \cite{Alvarez:2012ub}. 

\begin{figure}
\centering
\includegraphics[width=0.9\textwidth]{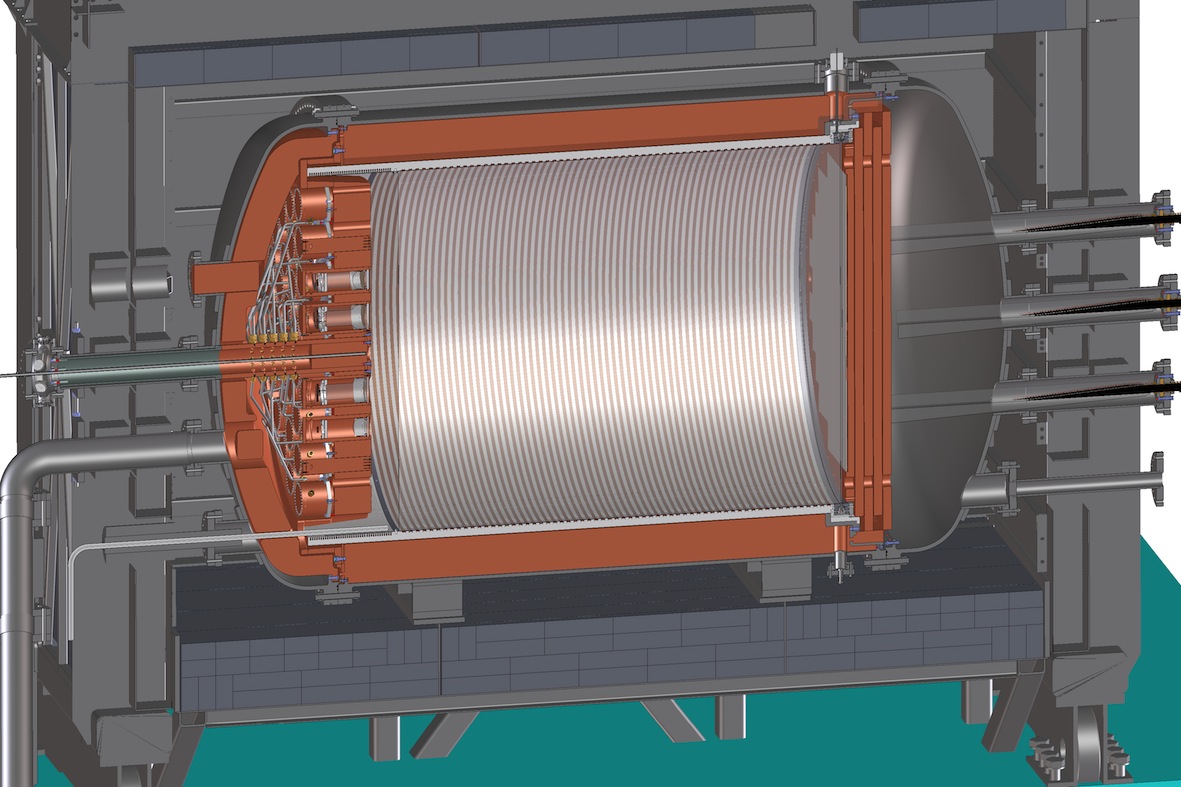}
\caption{Cross-section drawing of the NEXT-100 TPC inside its lead shielding.} \label{fig:Next100}
\end{figure}

The active volume of NEXT-100 --- a cylinder 130 cm long and 105 cm diameter --- can hold  100 kg of xenon gas at 15 bar, and it is surrounded by a copper shell 15 cm thick that shields it against external backgrounds. The TPC and the detector planes are housed in a stainless-steel (316Ti alloy) pressure vessel. A lead castle made of 20 cm thick bricks attenuates the external flux of high-energy gammas emanating from the laboratory walls. A drawing of the detector and shielding is shown in figure~\ref{fig:Next100}.

\section{Projected sensitivity of the NEXT-100 experiment} \label{sec:Next100}
NEXT has two powerful handles to distinguish signal from background:
\begin{itemize}
\item \emph{Energy resolution}: Signal events have all the same energy. Selecting only the events in the energy region around \Qbb\ defined by the resolution eliminates most of the spurious activity in the detector.
\item \emph{Event topology}: Electron tracks in high-pressure xenon (HPXe) are tortuous due to multiple scattering, and have a distinctive energy deposition pattern with a roughly-constant d$E/$d$x$ except at the end, where it increases by a factor of 4 to 5, leaving a large deposition that we call a \emph{blob}. This topological signature can be used in the search for \bbonu\ to distinguish between signal (two electron tracks with a common vertex) and background (mostly single electrons originated in the interaction of high-energy gammas), as demonstrated by the Gotthard experiment \cite{Luscher:1998sd}.\end{itemize}

The relevance of a background source depends therefore on its probability of generating a signal-like track in the active volume with energy around \Qbb. The \bbonu\ peak of \XE\ is located between the photo-electric peaks of the high-energy, de-excitation gammas emitted following the $\beta$-decays of \BI, from the uranium series, and \TL, from the thorium series. Most materials contain impurities of those isotopes in a given amount, and thus a thorough radiopurity control is being performed for the construction of the detector. Measurements of activity levels in the most relevant materials have been carried out based on GDMS and on ultra-low background germanium gamma-ray spectrometry at the LSC Radiopurity Service \cite{Alvarez:2012as}. 

Of special importance are the photomultipliers, which may be the dominant source of background in NEXT-100. Currently there are only upper limits to their background level. The most sensitive measurement, performed by the LUX collaboration, is and upper limit in the background of each PMT of less than 700 $\mu$Bq \cite{Malling:2011va}, while the XENON collaboration quotes a less sensitive limit of about 5 mBq/PMT \cite{Aprile:2011ru}. The NEXT Collaboration plans to screen the 60 PMTs to be installed in the detector, but for the moment we consider the quoted measurements as the optimistic and conservative limits, respectively, of the contribution of the PMTs to the total background rate.

\begin{figure}
\centering
\includegraphics[width=0.75\textwidth]{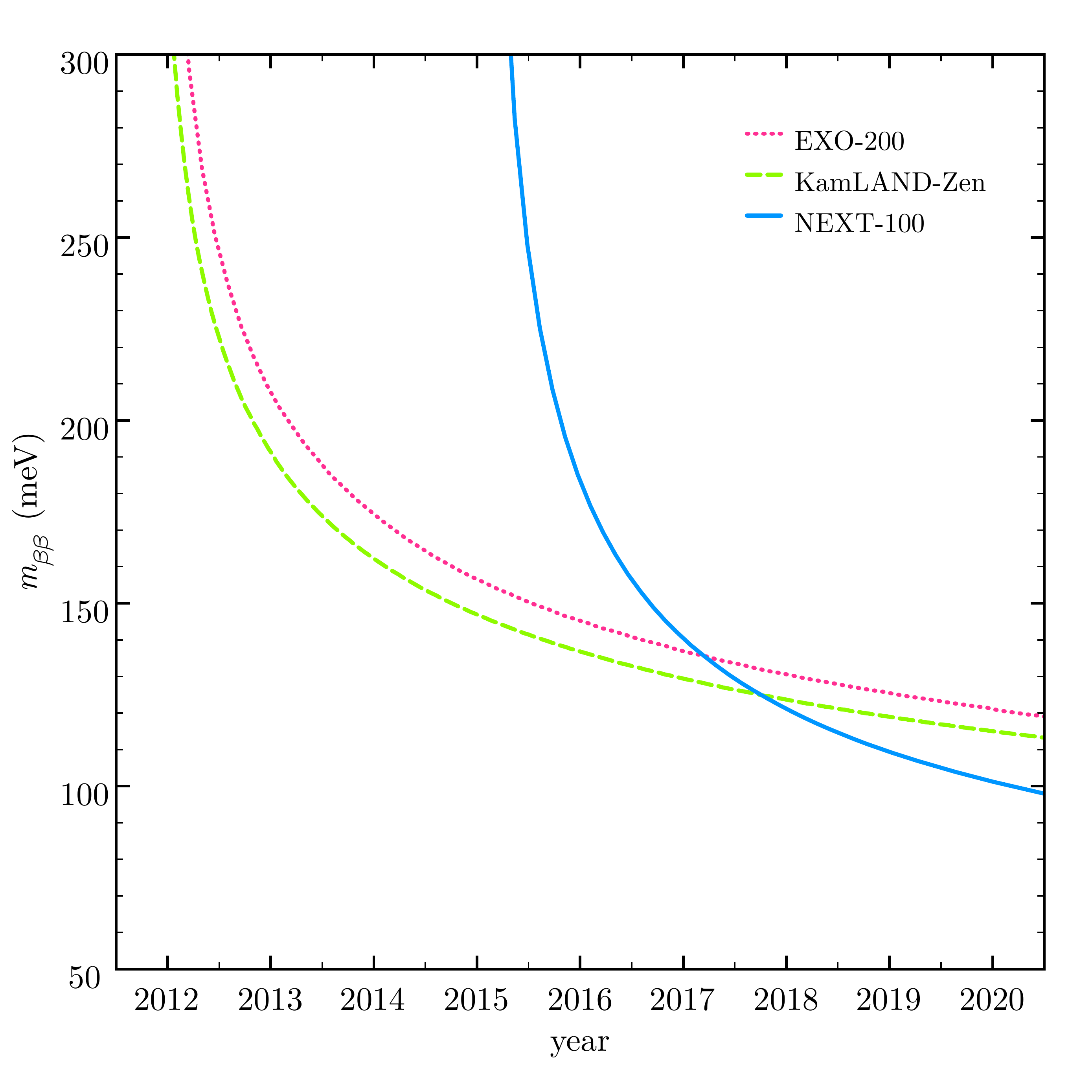}
\caption{Projected sensitivity of NEXT-100 (continuous, blue line) in a five-years run (2015--2020) compared with the sensitivities of the other two \XE-based experiments, KamLAND-Zen (dashed, green line) and EXO-200 (dotted, magenta line). The sensitivities have been calculated assuming the parameters in Table~\ref{tab:ExpParams} and a 10\% deadtime for the three experiments.}\label{fig:Sensitivity}
\end{figure}

Table~\ref{tab:Background} summarizes the contributions of the main subsystems of NEXT-100 to the total background rate, computed using a detailed simulation of the detector and the measurements from the radiopurity campaign. A conservative background rate of $5\times10^{-4}$~\ckky\ is obtained when the XENON limit is used for the background level of the PMTs, whereas a rate of about $10^{-4}$~\ckky\ is obtained for the LUX limit. Taking into account the conservative background estimate, and the computed  signal efficiency (see table~\ref{tab:ExpParams}), we have evaluated the sensitivity of the NEXT-100 experiment and compare it to that of KamLAND-Zen and EXO-200. The result is shown in figure~\ref{fig:Sensitivity}. The curve corresponding to the NEXT-100 experiment drops faster than those of the other two experiments, due to better energy resolution and background suppression. This compensates its late start. By 2018 the three experiments will reach a similar sensitivity of about 130~meV. By 2020, NEXT-100 will reach, after a 5-years run, a sensitivity of about 100 meV.

\begin{table}
\lineup
\centering
\caption{Estimated contribution of several subsystems to the total background rate of the NEXT-100 detector.} \label{tab:Background}
\begin{tabular}{lcc}
\br
Subsystem & \TL\ ($10^{-4}$~ckky) & \BI\ ($10^{-4}$~ckky) \\
\mr
Pressure vessel & 0.1 & 0.1 \\
Field cage 		& 0.2 & 0.2 \\
Energy plane 	& 2.2 & 0.9 \\
Tracking plane 	& 1.0 & 0.2 \\ 
\mr
Total			& 3.5 & 1.4 \\
\br
\end{tabular}
\end{table}

\section{Conclusions}
Neutrinoless double beta decay experiments are the only practical way to establish whether neutrinos are Majorana particles. The current generation of experiments, with a sensitivity to \mbb\ of about 100~meV, is already operating, with initial results from EXO-200 and KamLAND-Zen, both using \XE. The NEXT-100 detector could contribute in a decisive way to this exploration, in spite of a relatively late start.

At the same time, exploring the inverse hierarchy will require large masses (circa 1 ton) and a background level in the regime of very few counts per ton-year. Many of the current technologies may be eliminated either by lack of resolution or by the difficulties to extrapolate to large masses. In contrast, the NEXT concept which combines very good resolution, a topological signature and a clean detector, could lead the way in the near future.

\ack
This work was supported by the {\it Ministerio de Econom\'ia y Competitividad} of Spain under grants CONSOLIDER-Ingenio 2010 CSD2008-0037 (CUP) and FPA2009-13697-C04-04.

\section*{References}
\bibliographystyle{iopart-num}
\bibliography{references}

\providecommand{\newblock}{}
\begin{thebibliography}{10}
\expandafter\ifx\csname url\endcsname\relax
  \def\url#1{{\tt #1}}\fi
\expandafter\ifx\csname urlprefix\endcsname\relax\def\urlprefix{URL }\fi
\providecommand{\eprint}[2][]{\url{#2}}

\bibitem{GomezCadenas:2011it}
G\'omez-Cadenas J~J, Mart\'in-Albo J, Mezzetto M, Monrabal F and Sorel M 2012
  {\em Riv.\ Nuovo Cim.\/} {\bf 35} 29--98 (\textit{Preprint}
  \eprint{1109.5515})

\bibitem{Avignone:2007fu}
Avignone~III F~T, Elliott S~R and Engel J 2008 {\em Rev.\ Mod.\ Phys.\/} {\bf
  80} 481--516 (\textit{Preprint} \eprint{0708.1033})

\bibitem{Furry:1939qr}
Furry W~H 1939 {\em Phys.\ Rev.\/} {\bf 56} 1184--1193

\bibitem{Schechter:1980gr}
Schechter J and Valle J~W~F 1980 {\em Phys.\ Rev.\ D\/} {\bf 22} 2227

\bibitem{Beringer:1900zz}
Beringer J {\em et~al.\/} (Particle Data Group) 2012 {\em Phys.\ Rev.\ D\/}
  {\bf 86} 010001

\bibitem{GonzalezGarcia:2007ib}
Gonz\'alez-Garc\'ia M~C and Maltoni M 2008 {\em Phys.\ Rept.\/} {\bf 460}
  1--129 (\textit{Preprint} \eprint{0704.1800})

\bibitem{Fukugita:1986hr}
Fukugita M and Yanagida T 1986 {\em Phys.\ Lett.\ B\/} {\bf 174} 45

\bibitem{Davidson:2008bu}
Davidson S, Nardi E and Nir Y 2008 {\em Phys.Rept.\/} {\bf 466} 105--177
  (\textit{Preprint} \eprint{0802.2962})

\bibitem{GomezCadenas:2010gs}
G\'omez-Cadenas J~J, Mart\'in-Albo J, Sorel M, Ferrario P, Monrabal F {\em
  et~al.\/} 2011 {\em JCAP\/} {\bf 1106} 007 (\textit{Preprint}
  \eprint{1010.5112})

\bibitem{KlapdorKleingrothaus:2000sn}
Klapdor-Kleingrothaus H~V {\em et~al.\/} 2001 {\em Eur.\ Phys.\ J.\ A\/} {\bf
  12} 147--154 (\textit{Preprint} \eprint{hep-ph/0103062})

\bibitem{KlapdorKleingrothaus:2001ke}
Klapdor-Kleingrothaus H~V, Dietz A, Harney H~L and Krivosheina I~V 2001 {\em
  Mod.\ Phys.\ Lett.\ A\/} {\bf 16} 2409--2420 (\textit{Preprint}
  \eprint{hep-ph/0201231})

\bibitem{Auger:2012gs}
Auger M {\em et~al.\/} (EXO Collaboration) 2012 {\em JINST\/} {\bf 7} P05010
  (\textit{Preprint} \eprint{1202.2192})

\bibitem{Auger:2012ar}
Auger M {\em et~al.\/} (EXO Collaboration) 2012 {\em Phys.\ Rev.\ Lett.\/} {\bf
  109} 032505 (\textit{Preprint} \eprint{1205.5608})

\bibitem{KamLANDZen:2012aa}
Gando A {\em et~al.\/} (KamLAND-Zen Collaboration) 2012 {\em Phys.\ Rev. C\/}
  {\bf 85} 045504 (\textit{Preprint} \eprint{1201.4664})

\bibitem{Gando:2012zm}
Gando A {\em et~al.\/} (KamLAND-Zen Collaboration) 2012  (\textit{Preprint}
  \eprint{1211.3863})

\bibitem{McCowan:2010zz}
McCowan P and Barber R 2010 {\em Phys.\ Rev.\ C\/} {\bf 82} 024603

\bibitem{Abe:2009aa}
Abe S {\em et~al.\/} (KamLAND Collaboration) 2010 {\em Phys.\ Rev.\ C\/} {\bf
  81} 025807 (\textit{Preprint} \eprint{0907.0066})

\bibitem{NEXT:2012haa}
\'Alvarez V {\em et~al.\/} (NEXT Collaboration) 2012 {\em JINST\/} {\bf 7}
  T06001 (\textit{Preprint} \eprint{1202.0721})

\bibitem{Alvarez:2012np}
\'Alvarez V {\em et~al.\/} (NEXT Collaboration) 2012  (\textit{Preprint}
  \eprint{1211.4838})

\bibitem{Alvarez:2012hh}
\'Alvarez V {\em et~al.\/} (NEXT Collaboration) 2012 {\em Nucl.\ Instrum.\
  Meth.\ A\/} (\textit{Preprint} \eprint{1211.4474})

\bibitem{Oliveira:2013}
Oliveira C~A~B 2013 {\em {J.\ Phys.\ Conf.\ Ser.}\/} (\textit{Preprint}
  \eprint{1301.XXXX})

\bibitem{Nygren:2009zz}
Nygren D 2009 {\em Nucl.\ Instrum.\ Meth.\ A\/} {\bf 603} 337--348

\bibitem{Alvarez:2012ub}
\'Alvarez V {\em et~al.\/} (NEXT Collaboration) 2012 {\em JINST\/} {\bf 7}
  P02010 (\textit{Preprint} \eprint{1201.2018})

\bibitem{Luscher:1998sd}
Luscher R {\em et~al.\/} 1998 {\em Phys.\ Lett.\ B\/} {\bf 434} 407--414

\bibitem{Alvarez:2012as}
\'Alvarez V {\em et~al.\/} 2012 {\em JINST\/} (\textit{Preprint}
  \eprint{1211.3961})

\bibitem{Malling:2011va}
Malling D {\em et~al.\/} 2011  (\textit{Preprint} \eprint{1110.0103})

\bibitem{Aprile:2011ru}
Aprile E {\em et~al.\/} (XENON Collaboration) 2011 {\em Astropart.\ Phys.\/}
  {\bf 35} 43--49 (\textit{Preprint} \eprint{1103.5831})

\end{thebibliography}

\end{document}